\documentstyle[graphicx]{article}

\begin{document}
\title{Resonant inverse Compton scattering by secondary pulsar plasma}
\author{ Y.E.~Lyubarskii, S.A.~Petrova    }
\date{Institute of Radio Astronomy, Chervonopraporna St.4, Kharkov, 310002 Ukraine}

\maketitle
\begin{abstract}
We consider resonant inverse Compton scattering of thermal photons by
secondary particles above the pulsar polar gap. At neutron star
temperatures higher than $10^5$ K the process appears to be an essential
energy loss mechanism for the particles. The distribution function of the
secondary plasma particles is found to be strongly affected by the scattering.
It becomes two-humped implying the development of two-stream instability. The
resonantly upscattered Compton photons are found to gain energy of
1--10 MeV forming an additional component in the pulsar gamma-ray spectrum. The
corresponding gamma-ray flux is estimated as well.
\end{abstract}

\section{Introduction}
Rotation of a highly magnetized neutron star is known to induce a strong
electric field, which intensely accelerates charged particles. According to the
customary polar gap models (Ruderman \& Sutherland 1975, Arons \& Scharlemann
1979), the acceleration takes place near the neutron star surface above the polar
cap, the Lorentz-factor of the primary particles increasing up to $\sim 10^6$.
The particles move along the magnetic lines and emit curvature photons,
which initiate pair-production cascade. The first electron-positron
pairs created screen the accelerating electric field, so that at higher altitudes the
particle energy remains unaltered; the typical Lorentz-factors of the secondary
plasma are $\sim 10-10^4$.

Recent observations testify to thermal soft X-ray emission from some pulsars
indicating that the neutron stars can be rather hot, $T\sim 5\cdot 10^5$ K,
while the polar caps can have still higher temperatures scaling a few times
$10^6$ K (Cordova {\it et al.} 1989, Ogelman 1991, Halpern \& Holt 1992, Finley
{\it et al.} 1992, Halpern \& Ruderman 1993, Ogelman \& Finley 1993,
Ogelman {\it et al.} 1993, Yancopoulos {\it et al.} 1994, Ogelman 1995,
Greiveldinger {\it et al.} 1996). Such high temperatures of the neutron star surface
are also predicted theoretically (Alpar {\it et al.} 1984, Shibazaki \& Lamb
1989, Van Riper 1991, Page \& Applegate 1992, Umeda {\it et al.} 1993, Halpern
\& Ruderman 1993). The thermal X-ray photons should suffer inverse Compton
scattering off the primary particles in the polar gap. In the presence of a
strong magnetic field the scattering cross-section is essentially enhanced,
if the photon energy in the particle rest frame equals the cyclotron energy
(Herold 1979, Xia {\it et al.} 1985). For the pulsars with hot polar caps the
resonant Compton scattering in the polar gap was found to be rather efficient
(Kardashev {\it et al.} 1984, Xia {\it et al.} 1985, Daugherty \& Harding
1989, Dermer 1990,
Sturner 1995, Chang 1995). Firstly, it was recognized as an essential mechanism
for energy loss of primary particles accelerating in the polar gap.
Secondly, inverse Compton scattering was found to condition the gap formation,
since the pair production avalanche may be triggered by the upscattered
Compton photons rather than by curvature photons (Zhang \& Qiao 1996, Qiao
\& Zhang 1996, Luo 1996, Zhang {\it et al.} 1997).

As shown by Sturner (1995), given typical values of neutron star temperature
and surface magnetic field, the resonant Compton scattering is the
strongest for particle Lorentz-factors $\sim 10^2-10^3$. Theoretical models
(Van Riper 1991, Page \& Applegate 1992, Umeda {\it et al.} 1993) suggest
that the typical temperatures of the neutron star surface are as high as a few
times $10^5$ K. Hence, the scattering is likely to be essential for the
secondary plasma in most of the pulsars. Daugherty \&
Harding (1989), Zhang {\it et al.} (1997)
traced the evolution of the Lorentz-factor of secondary particles
on account of magnetic inverse Compton scattering above the polar gap. Our
aim is to investigate in more detail the influence of resonant inverse
Compton scattering on the parameters of secondary plasma. For the resonant
character of the scattering, the evolution of particle Lorentz-factor with
the distance depends strongly on the initial particle energy. Since the
distribution function of the secondary plasma is generally believed to be
rather broad (the Lorentz-factor ranges from $10$ to $10^4$), its evolution
on account of the Compton scattering is of a great interest. It will be
shown that only particles with Lorentz-factors between 100 and 1000 are
essentially decelerated in the course of
the resonant scattering forming a sharp peak at low energies. Particles with
larger Lorentz-factors are not decelerated at all. Thus,
the resultant distribution function of secondary particles becomes
two-humped, giving rise to the two-stream instability.

In Sect. 2 we examine how the resonant inverse Compton scattering affects the
Lorentz-factor of a secondary particle at various pulsar parameters. Section 3
is devoted to studying the evolution of the distribution function. The
conditions for the development of the two-stream instability are also
discussed. In Sect. 4 we estimate the gamma-ray luminosity caused by
the upscattered Compton photons. Section 5 contains a brief summary.

\section{Deceleration of secondary particles due to resonant inverse Compton
scattering}
Consider the flow of secondary plasma streaming along the open magnetic field
lines above the polar gap. The neutron star is supposed to emit a blackbody
radiation, which is scattered by the plasma particles. In a strong magnetic
field the scattering is particularly efficient, if the photon energy,
$\varepsilon mc^2$, satisfies the resonance condition:
$$
\varepsilon\gamma (1-\beta\cos\theta)=\varepsilon_B, \eqno (1)$$
where $\gamma$ is the Lorentz-factor of the scattering particles, $\beta$
the particle velocity in units of $c$, $\theta$ the angle the photon makes
with the particle velocity, $\varepsilon\equiv B/B_{cr}$, with $B_{cr}=
m^2 c^3/\hbar e=4.414\cdot 10^{13}$ G.
At the distance $z$ from the neutron star the rate of energy
loss due to the resonant inverse Compton scattering can be written as
(Sturner 1995):
$$
\frac{d\gamma}{dt}=4.92\cdot 10^{11}\frac{T_6B_{12}^2(z)}{\beta\gamma}
$$$$
\times\ln\left[1-\exp\left(-\frac{\varepsilon_Bmc^2}
{\gamma (1-\beta\cos\theta_c(z))kT}\right)\right]{\rm s}^{-1}.
\eqno (2)$$
Here $T_6$ is the neutron star temperature in units of $10^6$ K, $B_{12}(z)$
the magnetic field strength in units of $10^{12}$ G, $\theta_c(z)$ the maximum
incident angle of photons,
$$
\cos\theta_c=\sqrt{1-\frac{1}{(1+z/R)^2}}, \eqno (3)$$
where $R$ is the neutron star radius. Provided that the magnetic field is
dipolar, $B_{12}(z)\propto (1+z/R)^{-3}$.
                                                       
Numerical solutions of Eq. (2) are presented in Fig. 1.

 One can see that the
resonant inverse Compton scattering is significant up to $z\sim R$ and it
affects the particle Lorentz-factor essentially. Note that the curves for
different initial Lorentz-factors are not similar to each other. In agreement
with Eq. (1), the larger the particle Lorentz-factor the lower is the energy
of resonantly scattered photons. At $\gamma_0=3000$ the energy of the resonant
photons is below $\varepsilon_{max}=2.82kT/(mc^2)$, which corresponds to the
maximum in the photon distribution. As $\gamma$ is decreasing with the altitude,
the resonant energy increases. Hence, the photon spectral density increases
and the scattering becomes more efficient. However, at distances $z\sim R$ it
ceases due to the essential shrinking of the solid angle subtended by the
neutron star. If the initial Lorentz-factor is $100$, the energy of resonant
photons is above $\varepsilon_{max}$ and further increases with the altitude,
so that the scattering gradually ceases.

Theoretical models predict that the polar cap region should be significantly
hotter than the rest surface of the neutron star ({\it e.g.} Cheng \& Ruderman 1980,
Arons 1981, Alpar {\it et al.} 1984, Umeda {\it et al.} 1993, Luo 1996).
However, the luminosities observed from the hot spots appear to be too small
(Finley {\it et al.} 1992, Halpern \& Ruderman 1993, Becker \& Trumper 1993,
Yancopoulos {\it et al.} 1994, Greiveldinger {\it et al.} 1996).
The latter is usually interpreted as a consequence of small hot spot radii
({\it e.g.} Yancopoulos {\it et al.} 1994, Greiveldinger {\it et al.} 1996).
As is evident from Fig.1, for typical hot spot parameters
the particle energy loss is mainly determined by the scattering
of the photons from the whole neutron star rather than by the scattering
of hot spot photons.
So hereafter we take into account only the photons from the whole neutron star
surface keeping in mind that the influence of hot spot photons can
somehow alter our quantitative results, while the qualitative picture should
remain the same.

It should be pointed out that the evolution of Lorentz-factor with the
altitude shown in Fig. 1 is somewhat different from that reported by
Zhang {\it et al.}(1997)(see their Fig. 1). These authors claim that
at distances $z\sim{\rm few}  R$ the particles suffer severe nonresonant
magnetized scattering, which leads to the drastic decrease of the
final Lorentz-factor. In fact, the rate of energy loss on account of
nonresonant magnetic scattering increases with decreasing the field strength
as $B_{12}^{-2}(z)$ (see Eq. (5) in Sturner 1995). However, this equation is
applicable only if in the particle rest frame the cyclotron energy exceeds
the energy of most of photons, $\varepsilon_B >\varepsilon\gamma (1-\beta\cos
\theta_c)$. For the photon energies at the peak of the Planck distribution,
$\varepsilon\approx 2.82kT$, the latter condition can be rewritten as
$4\cdot 10^{-2}B_{12}(z)/(T_6\gamma_3(1-\beta\cos\theta_c))>1$, with
$\gamma_3\equiv\gamma/10^3$. Taking $T_6=0.5$, $\gamma_3=3$, $B_{12}=0.1$
(Zhang {\it et al.} 1997, Fig. 1a), one can obtain that at $z=10R$ the
left-hand side of this inequality is $5\cdot 10^{-5}$, while at $z=0$ it
equals $3\cdot 10^{-4}$. So even at the stellar surface most of the photons
scatter in the Thomson regime rather than in the magnetic one. The rate of
energy loss out of the Thomson scattering is found to be
$$
\frac{d\gamma_T}{dt}=-30\gamma T_6^4(1-\beta\cos\theta_c)^3\,{\rm s}^{-1},
\eqno (4)$$
showing that this process is inefficient. Thus the resonant inverse Compton
scattering is the only significant energy loss mechanism for the secondary
particles in pulsar magnetospheres.

We believe that the particles start deceleration just above the polar
gap. In general the gap thickness is supposed to be of the order of the polar
cap radius (Ruderman \& Sutherland 1975) or even larger (Arons \& Scharlemann
1979). The energy loss of secondary particles due to the resonant
inverse Compton scattering above the gap should certainly be influenced by
the adopted value of the gap height. In Fig. 2 we show the final
Lorentz-factors versus the gap height. One can see that the dependence is
sufficiently weak, therefore, below we fix $z_0$ at $10^{-2}R$.

In contrast with the gap height, such pulsar parameters as the star temperature
and magnetic field strength can influence particle deceleration essentially.
In Fig. 3 we present the dependences of the normalized final Lorentz-factor
on the neutron star temperature at various magnetic field strengths.
Apparently, at the temperatures $<10^5$ K the scattering is inefficient yet,
while at higher temperatures it becomes significant. Note that at various
initial Lorentz-factors ($\gamma_0=100$ and $3000$) the scattering efficiency
versus magnetic field strength is essentially different. The particles with
$\gamma_0=100$ resonantly scatter the photons from the Wien region. Then the
weaker the field strength, the lower is the energy of resonant photons and,
correspondingly, the higher is the photon spectral density and the larger
is the particle energy loss (see Fig. 3a). At $\gamma_0=3000$ the resonant
energy lies in the Rayleigh-Jeans region. So the scattering is more efficient
for higher magnetic field strengths, since the photon spectral density
increases with the energy (see Fig. 3b).

For the resonant character of the scattering the particle energy loss
depends strongly on the initial Lorentz-factor. Figure 4 shows the final
Lorentz-factor versus the initial one for various neutron star temperatures
and magnetic field strengths. At $\gamma_0\sim 10$ as well as at $\gamma_0
\sim 10^4$ the resonant scattering appears to be inefficient. Provided that the
scattering is intense ($\gamma_0\sim 10^2-10^3$), the final Lorentz-factor
appears to be independent of the initial one. According to Fig. 4a, the length
of the plateau increases with the star temperature. In fact, the higher
temperature implies the larger amount of photons at every energy, the
scattering becoming more efficient. As can be seen from Fig. 4b, the increase
of the magnetic field strength leads to the shift of the plateau toward a
higher energy; this is certainly consistent with the resonance condition (1).

\section{The distribution function of the secondary plasma as a result of
resonant inverse Compton scattering}
Since the energy loss depends essentially on the initial particle energy,
we are to investigate  the evolution of the distribution function of
secondary particles as a result of the resonant inverse Compton scattering.
Conservation of the number of particles along a phase trajectory
implies that
$$
f(z,\,\gamma)d\gamma=f(z_0,\,\gamma_0)d\gamma_0\,,$$
where $f(z,\,\gamma)$ is the particle distribution function and the
subscript ''0'' refers to the initial values. So looking for the phase
trajectories of individual particles one can reconstruct the distribution
function at any height $z$. We are particularly interested in the final
distribution function arising at distances, where the resonant scattering
ceases.

Let us begin with the evolution of a waterbag distribution function:
$$
f(z_0,\,\gamma_0)=
\left\{
\begin{array}{l}
\frac{1}{\gamma_{max}-\gamma_{min}},\quad\qquad\qquad \gamma_{min}\leq\gamma\leq
\gamma_{max},\\
0\,,\,\quad\qquad\qquad\qquad \gamma <\gamma_{min}\,{\rm  and }\,\gamma >\gamma_{max},\\
\end{array}
\right.$$
with $\gamma_{min}=10$, $\gamma_{max}=10^4$.
The final distribution functions
at various pulsar
parameters are plotted in Fig. 5. The particles with $\gamma\sim 10^2-10^3$
are essentially decelerated due to the scattering, the final energies becoming
equal (see also Fig. 4). These particles form the sharp peak at $\gamma
<10^2$. For the Lorentz-factors of a few thousand the scattering becomes
inefficient and the distribution function remains almost unaltered. Thus,
the resonant inverse Compton scattering leads to the two-humped distribution
function of the secondary plasma. At higher neutron star temperatures the
main peak of the function shifts towards the lower energies and the humps
become more prominent. The magnetic field strength variation also results
in the shift of the main peak. For the more realistic distribution function
resembling that found by Arons (1980),
$$
f(\gamma)=
\left\{
\begin{array}{l}
\exp\left(-10\frac{\gamma_m-\gamma}{\gamma_m}\right),\quad\qquad\qquad 10\leq\gamma\leq
\gamma_m,\\
(\gamma/\gamma_m)^{-3/2},\,\quad\qquad\qquad\qquad \gamma_m\leq\gamma\leq\gamma_c,\\
(\gamma_c/\gamma_m)^{-3/2}\exp\left(-\frac{\gamma-\gamma_c}{\gamma_c}\right),
\quad \gamma_c\leq\gamma\leq 10^4,\\
\end{array}
\right.\eqno (5)$$
with $\gamma_m=10^2$, $\gamma_c=10^{3.5}$, the evolution on account of
the resonant inverse Compton scattering is qualitatively the same (see Fig. 6).

We next examine the dispersion properties of the plasma with the evolved
distribution function. For simplicity let us assume that the plasma consists
of the two particle flows characterized by the number densities $n_a$,
$n_b$ and by the Lorentz-factors $\gamma_a$, $\gamma_b$ ($\gamma_a <\gamma_b$).
Given the infinitely strong magnetic field, the dispersion relation is as
follows (see, {\it e.g.} Lyubarskii 1995):
$$
\frac{\omega_{pa}^2}{(\omega -kv_a)^2\gamma_a^3}+\frac{\omega_{pb}^2}
{(\omega -kv_b)^2\gamma_b^3}=1. \eqno (6)$$
Here $\omega$ is the frequency, $k$ the wave number, $v_{a,b}$ are the
particle velocities, $\omega_{pa,b}$ the plasma frequencies given by the
customary expression:
$$
\omega_{pa,b}=\sqrt{\frac{4\pi n_{a,b}e^2}{m}}, \eqno (7)$$
where $e$ is the electron charge, $m$ the electron mass. As is evident from
Eq. (6), the dispersion properties of the plasma are mainly determined by one
of the particle flows on condition that
$$
\frac{n_a}{\gamma_a^3}\gg\frac{n_b}{\gamma_b^3}, \eqno (8)$$
rather than $n_a\gg n_b$. This is because of the great inertia of the
fast particles performing one-dimensional motion in the superstrong magnetic
field. For the distribution functions plotted in Fig. 5 and 6 $n_a$ and
$n_b$ are comparable, while $\gamma_a/\gamma_b\sim 10^{-1}-10^{-2}$. Therefore
the low-energy particles of the main peak almost completely determine the
dispersion properties of the plasma.

The two-humped distribution function implies the possibility of the
two-stream instability. Provided that the contribution of one of the plasma
flows to the plasma dispersion is small ({\it i.e.} Eq. (8) is valid), the
growth rate of the instability takes the form (Lominadze \& Mikhailovskii 1979,
Cheng \& Ruderman 1980):
$$
{\rm Im} \omega\approx\left(\frac{n_b}{n_a}\right)^{1/3}\frac{\omega_{pa}}
{\gamma_a^{1/2}\gamma_b}. \eqno (9)$$
The two-stream instability results in an essential development of
initial perturbations on condition that
$$
\frac{R_c}{c}{\rm Im}\omega> 10, \eqno (10)$$
where $R_c$ is the characteristic scale length for the increase of
perturbations.

It is convenient to normalize the number density of the secondary plasma by
the Goldreich-Julian charge density:
$$
n=\frac{\kappa B}{Pce}, \eqno (11)$$
where $\kappa$ is the multiplicity factor of the secondary plasma, $P$ the
pulsar period. The
latter equation can be rewritten as:
$$
n=6.25\cdot 10^{13} P^{-1}\kappa_3B_{12}(1+z/R)^{-3}\, {\rm cm^{-3}},\eqno (12)$$
where $\kappa_3\equiv\kappa/10^3$. Using Eqs. (9) and (12) in Eq. (10) we
reduce the condition for the efficient instability development to the form:
$$
\frac{\kappa_3B_{12}}{PR_{c7}\gamma_{b3}^2\gamma_{a2}}>4\cdot 10^{-4}.
\eqno (13)$$
Here $R_{c7}\equiv R_c/10^7{\rm cm}$, $\gamma_{b3}\equiv \gamma_b/10^3$,
$\gamma_{a2}\equiv\gamma_a/10^2$ and it is assumed that $(n_b/n_a)^{1/3}
\approx 1$. As can be seen from Eq. (13), at typical pulsar parameters the
two-stream instability can develop readily providing the increase of plasma
oscillations. The latter can be transformed into electromagnetic waves,
thus giving rise to pulsar radio emission.

It should be mentioned that two-steram instability has always been one of
the most popular mechanisms for pulsar radio emission.
For more than two
decades a number of scenarios for the instability development were proposed.
The first and the most natural one involves the instability caused by
the flows of primary and secondary pulsar plasma (Ruderman \& Sutherland 1975).
However, the growth rate of this instability appears to be too small
because of enormous inertia of the high-energy primary particles (Benford \&
Buschauer 1977). Cheng \& Ruderman (1977) considered the two-stream
instability arising in the secondary plasma due to the difference in the
velocities of electrons and positrons moving along the curved magnetic lines.
However, this difference is insufficient to cause the instability, since the
particle distribution functions are too broad (Buschauer \& Benford 1977).
Lyubarskii (1993) suggested that current  and charge density adjustment
in pulsar
magnetosphere leads to the backward particle flow, which
causes intense two-stream instability. However, numerical simulations of the
plasma flow in the open field line tube are necessary to prove this idea.
Given nonstationary generation of the secondary plasma the particles are
confined to the separate clouds, and the fastest particles of a cloud can
outstrip the slower particles of the previous cloud giving rise to the
two-stream instability (Usov 1987, Ursov \& Usov 1988). Up to now it is not
clear whether the instability initiated in such a way can account for pulsar
radio emission, since the nonstationarity of plasma generation is not studied
in detail yet.
The present
paper suggests one more possibility of two-stream instability in pulsars,
which is based on the selective energy loss of particles as a result of
resonant inverse Compton scattering. Note that in this model the instability
arises naturally, with no additional assumptions being involved.

\section{Gamma-ray luminosity provided by resonantly
upscattered Compton photons}
The photons produced by the resonant inverse Compton scattering have
the energies
$$
E\sim\gamma\epsilon_Bmc^2\sim B_{12}\gamma_2\,{\rm MeV},\eqno (14)$$
so that typically $E\sim 1-10$ MeV. Note that the curvature gamma-photons
produced in the polar gap as well as the photons upscattered by the primary
particles have essentially higher energies, $E\sim 100$ MeV. Hence, the resonant
scattering by secondary plasma results in an additional low-energy component in pulsar
gamma-ray spectrum. The spectrum of this component in the case of some
specific distribution functions of the scattering particles is obtained
by Daugherty \& Harding (1989).
In general, the low-energy tail of this component spreads even to the
X-ray band on account of the photons resonantly scattered at distances
$z>R$, where the magnetic field strength decreases essentially. However,
these photons are very few, since at $z>R$ the scattering rate decreases
significantly due to the shrinking of the solid angle subtended by the
neutron star.

Given that the scattering is efficient, most of the energy of the secondary
plasma  should be transferred to the low-energy gamma-rays. The luminosity
provided by the upscattered photons can be estimated as follows:
$$
L=nSmc^3\Delta\gamma,\eqno (15)$$
where $S$ is the cross-sectional area of the open field line tube,
$$
S=\frac{\pi R^3}{R_L},
\eqno (16)$$
$R_L$ is the light cylinder radius, $\Delta\gamma$ the difference between
the Lorentz-factors of the
particles, which mainly contribute to the final and initial plasma energy,
$\Delta\gamma\sim 10^2-10^3$. Substituting Eqs. (12) and (16) into Eq. (15)
we find:
$$
L=0.942\cdot 10^{30}\frac{B_{12}R_6^3\kappa_3\gamma_3}{P^2}\, {\rm ergs/s}.
\eqno (17)$$

Although in the particle rest frame the photons are scattered in all
directions, in the laboratory frame they are beamed along the particle velocity.
So the opening angle of the gamma-ray beam is given by
$\varphi =3\sqrt{R/R_L}.$
The averaged observed photon flux, $F$, is related to the luminosity as
$$
F=\frac{L\varphi}{2\pi\Omega d^2E\Delta E},\eqno (18)$$
where $\Omega =\pi\varphi^2/4$ is the solid angle occupied by the beam of
upscattered photons,
$d$ the distance to the pulsar, $\Delta E$ the energy band. Taking into
account Eq. (17), Eq. (18) can be rewritten as
$$
F=3\cdot 10^{-10}\frac{B_{12}R_6^{5/2}\gamma_3\kappa_3}{P^{3/2}d_3^2 E_6^2}\,
{\rm photons/(cm^2\cdot s\cdot keV)}, \eqno (19)$$
where $d_3\equiv d/10^3\,{\rm pc}$, $E_6\equiv E/1\,{\rm MeV}$.

For most pulsars the flux given by Eq. (19) is too low to be detected.
At present the detectors of the Compton gamma-ray observatory are the most
sensitive to the low-energy gamma-ray emission (OSSE at 0.05--10 MeV and
COMPTEL at 1--30 MeV). At 1 MeV the source sensitivity of OSSE is only
$2\cdot 10^{-7}$ photons/(${\rm cm^2\cdot s\cdot keV}$) (Gehrels \&
Shrader 1996). In the observations reported by Kuiper {\it et al.} (1996) the
flux from Geminga in the band of 3--10 MeV was found to be $10^{-7}E_6^{-2}$
photons/(${\rm cm^2\cdot s\cdot keV}$). Substituting Geminga parameters
($P=0.237$ s, $B_{12}=3.3$, $d_3=0.15$) into Eq. (19) one can obtain the flux
provided by the upscattered Compton photons: $F=3.8\cdot 10^{-7}\kappa_3
\gamma_3R_6^{5/2}E_6^{-2}$ photons/(${\rm cm^2\cdot s\cdot keV}$), which is
consistent with the one detected.

\section{Conclusions}
We have investigated resonant inverse Compton scattering by secondary
pulsar plasma. The process is found to cause the efficient energy loss of
the secondary particles given the neutron star temperatures $>10^5$ K, so
that our results are applicable to most pulsars. For the resonant
character of the scattering, the energy loss depends strongly on the initial
particle energy. At $\gamma_0\sim 10^2-10^3$ the scattering is the most
essential, the final Lorentz-factors of the particles being independent of
the initial ones.
                                                                  
The distribution function of the secondary plasma is significantly altered by
the resonant inverse Compton scattering. It is shown that ultimately the
distribution function becomes two-humped. The main peak at $\gamma\sim 10^2$
is very sharp. It is formed by particles which suffered severe energy
loss on account of the scattering. Another hump is sufficiently broad. It is
associated with the particles, whose Lorentz-factors are almost unaltered by
the scattering. The two-humped distribution function of the plasma particles is
known to be unstable. It is shown that at pulsar conditions the two-stream
instability develops readily and leads to an essential increase of plasma
oscillations, which are likely to be transformed into radio emission.

We have also estimated the gamma-ray flux provided by the upscattered Compton
photons. The resonantly scattered photons appear to gain the energies of
1--10 MeV forming an additional low-energy component in pulsar gamma-ray spectrum.

\begin{figure}[htb]
\includegraphics[scale=0.4]{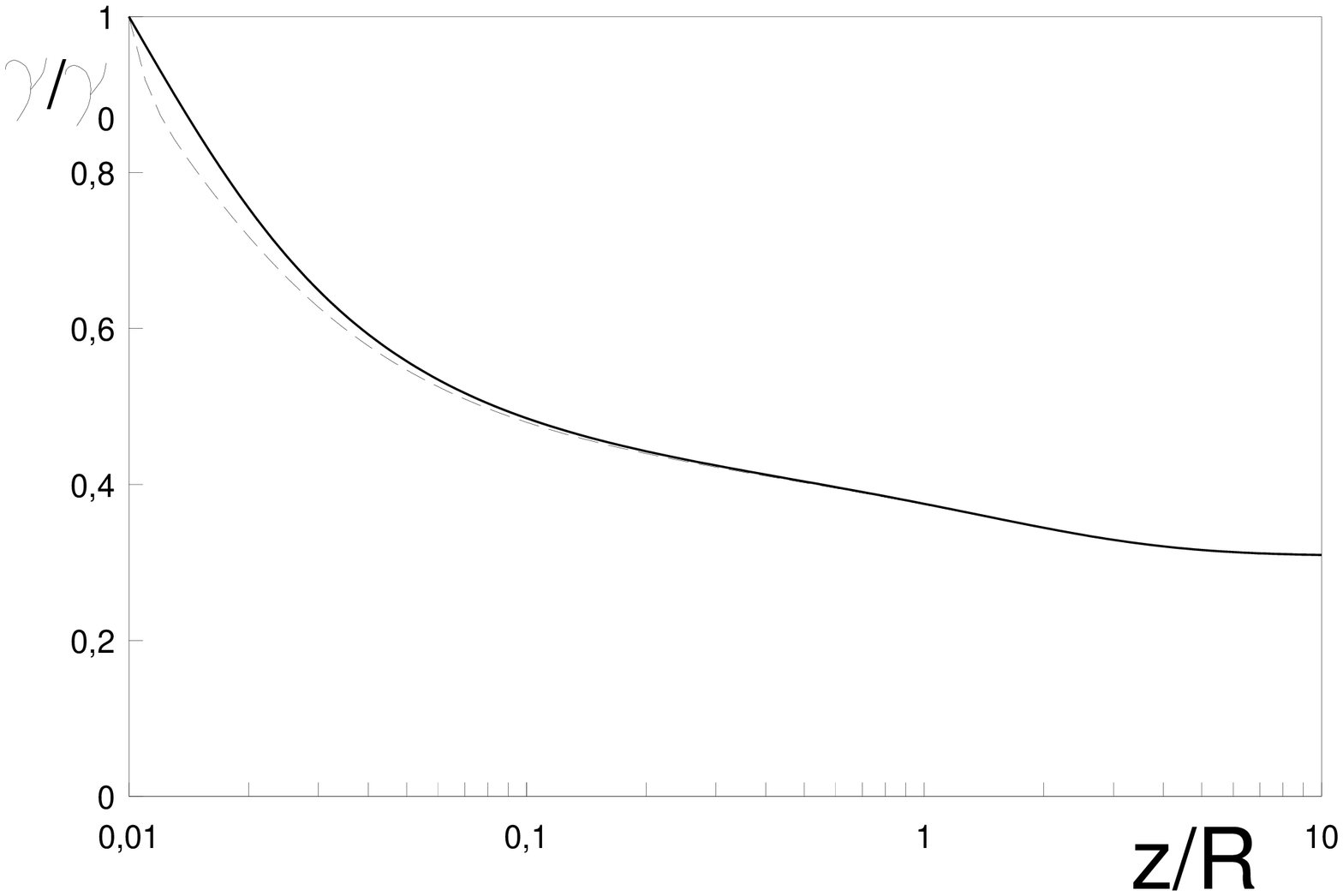}
\includegraphics[scale=0.4]{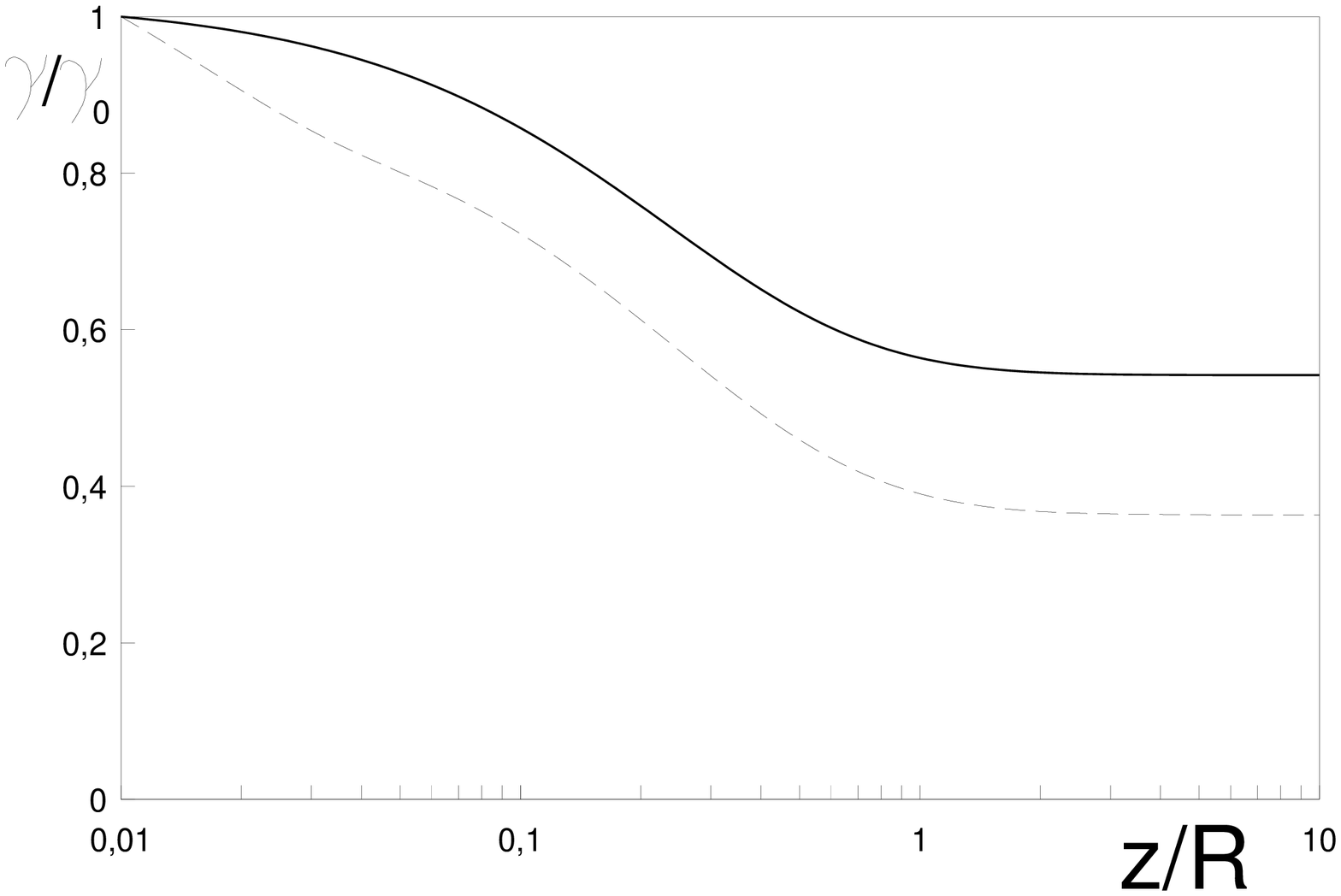}
\caption{
 Evolution of particle Lorentz-factor with the distance for
 initial Lorentz-factors, $\gamma_0=100$ (a) and $\gamma_0=3000$ (b);
 $B_{12}=1$, $z_0/R=0.01$.The solid lines are plotted for the case
of scattering the photons from the whole neutron star surface with the
temperature $5\cdot 10^5$K, whereas the dashed lines refer to the case when the
contribution of the hot spot photons is also taken into account(here the
hot spot radius is $5\cdot 10^{-3}R$ and the temperature is $3\cdot 10^6$K).
}
\end{figure}

\begin{figure}[htb]
\includegraphics[width=\textwidth,keepaspectratio]{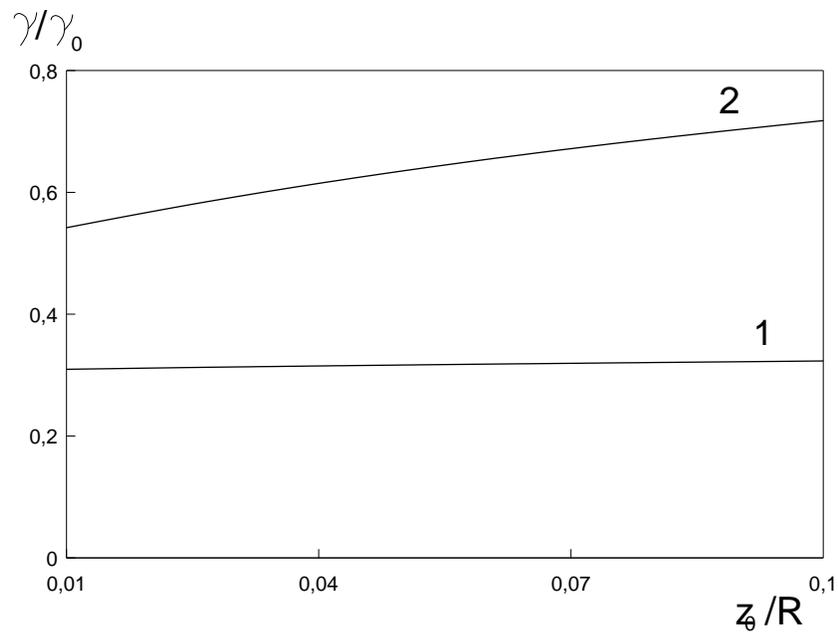}
\caption{
 Final Lorentz-factor versus the gap height for
$\gamma_0=100$ (curve 1) and $\gamma_0=3000$ (curve 2); $T_6=0.5$, $B_{12}=1$ 
}
\end{figure}

\begin{figure}[htb]
\includegraphics[scale=0.4]{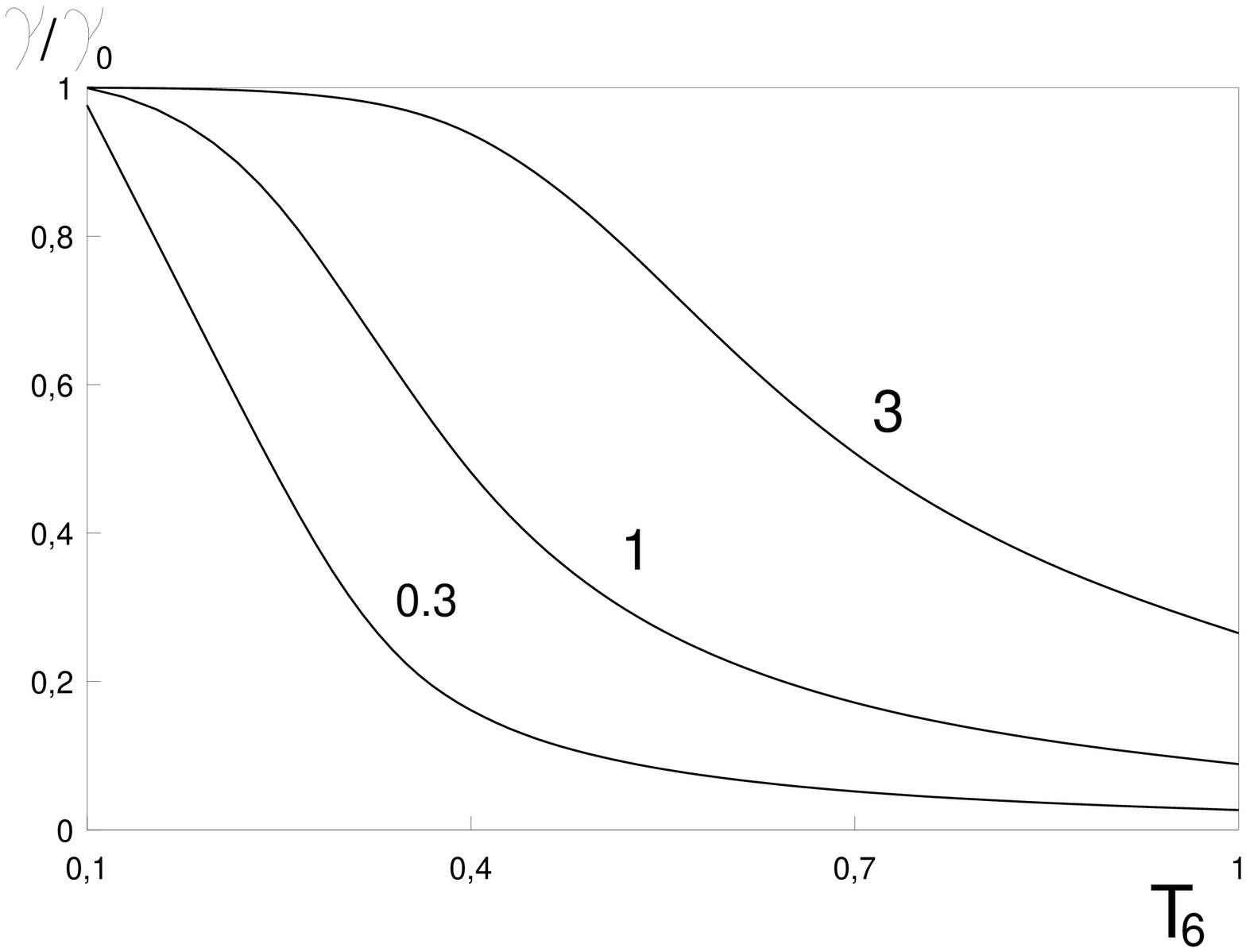}
\includegraphics[scale=0.4]{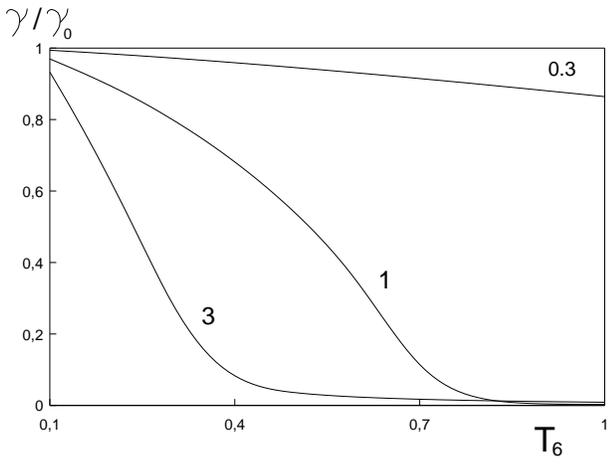}
\caption{
Final Lorentz-factor versus the temperature for various
magnetic field strengths, $B_{12}$: {\bf a} $\gamma_0=100$, {\bf b}
$\gamma_0=3000$
}
\end{figure}

\begin{figure}[htb]
\includegraphics[scale=0.4]{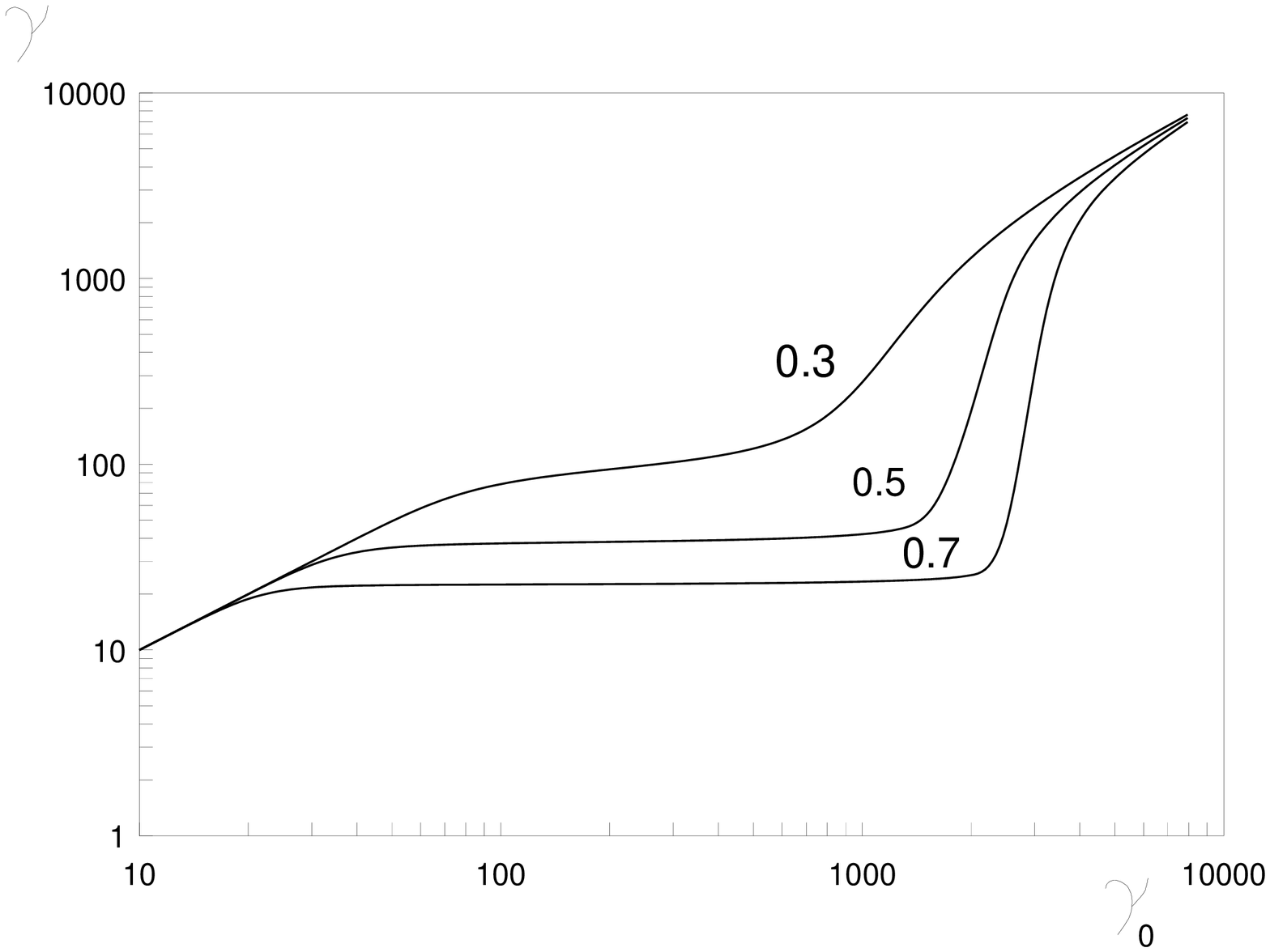}
\includegraphics[scale=0.4]{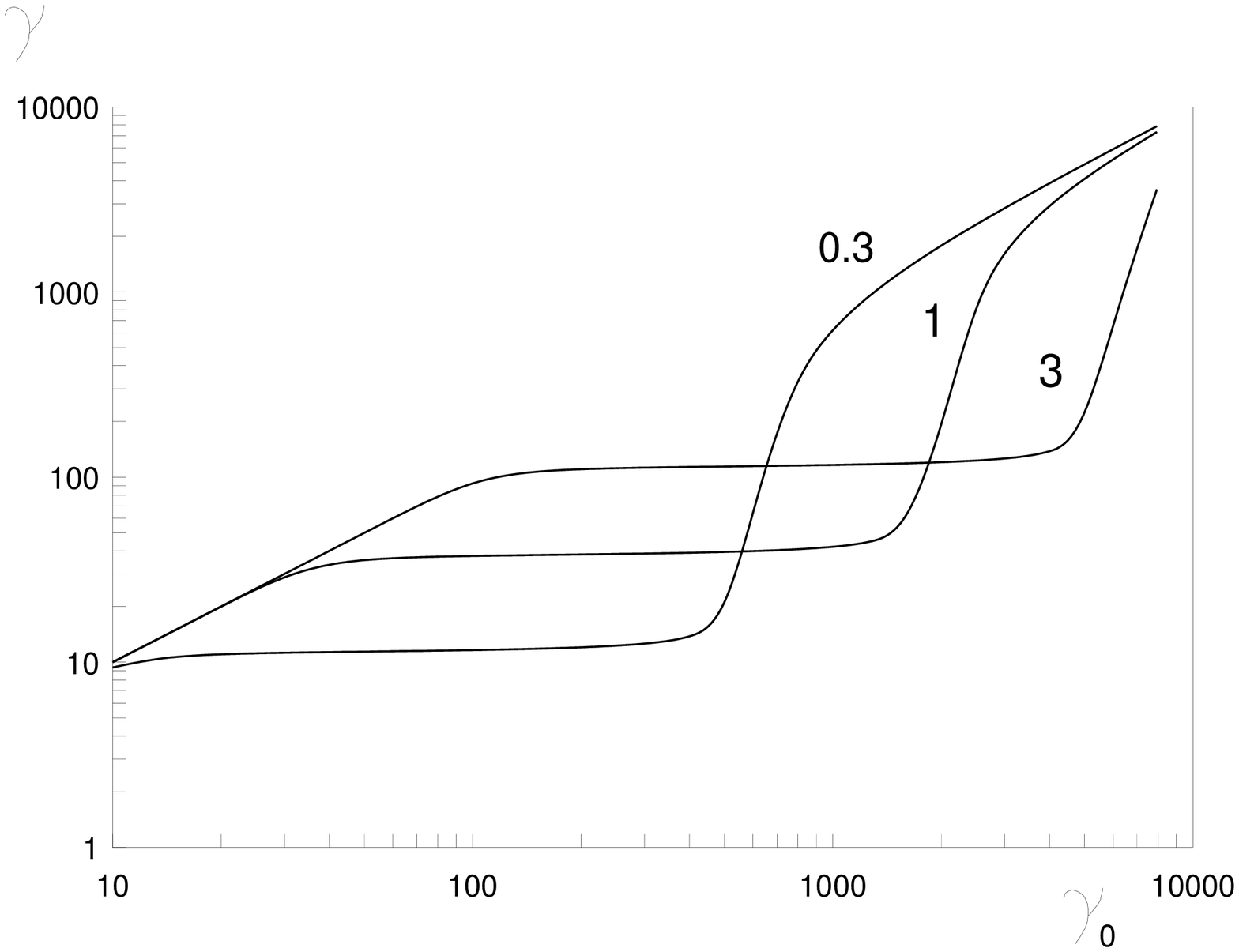}
\caption{
 Final Lorentz-factor versus the initial one: {\bf a}
for various temperatures, $T_6$; $B_{12}=1$, {\bf b} for various
magnetic field strengths, $B_{12}$; $T_6=0.5$
}
\end{figure}

\begin{figure}[htb]
\includegraphics[scale=0.4]{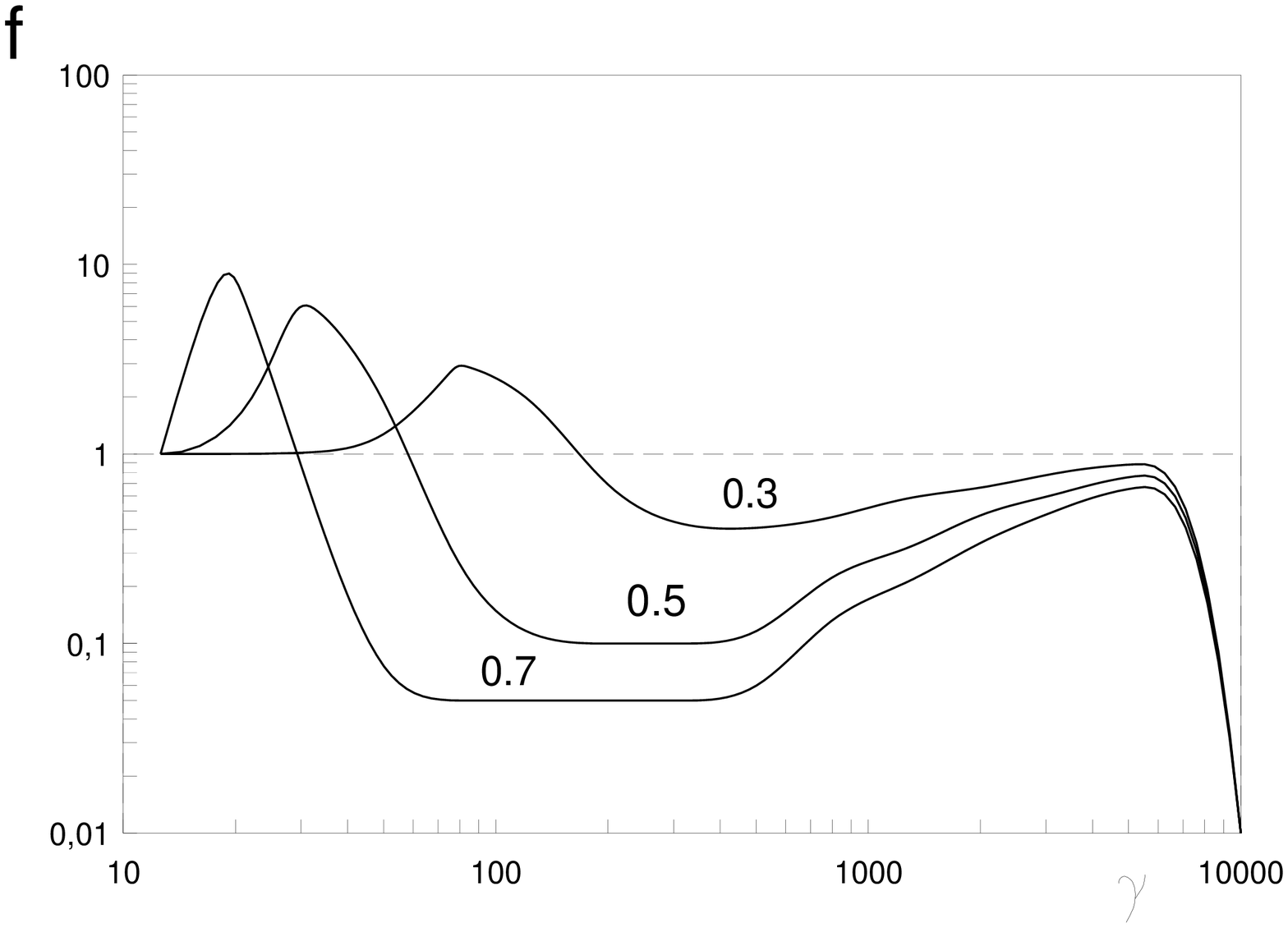}
\includegraphics[scale=0.4]{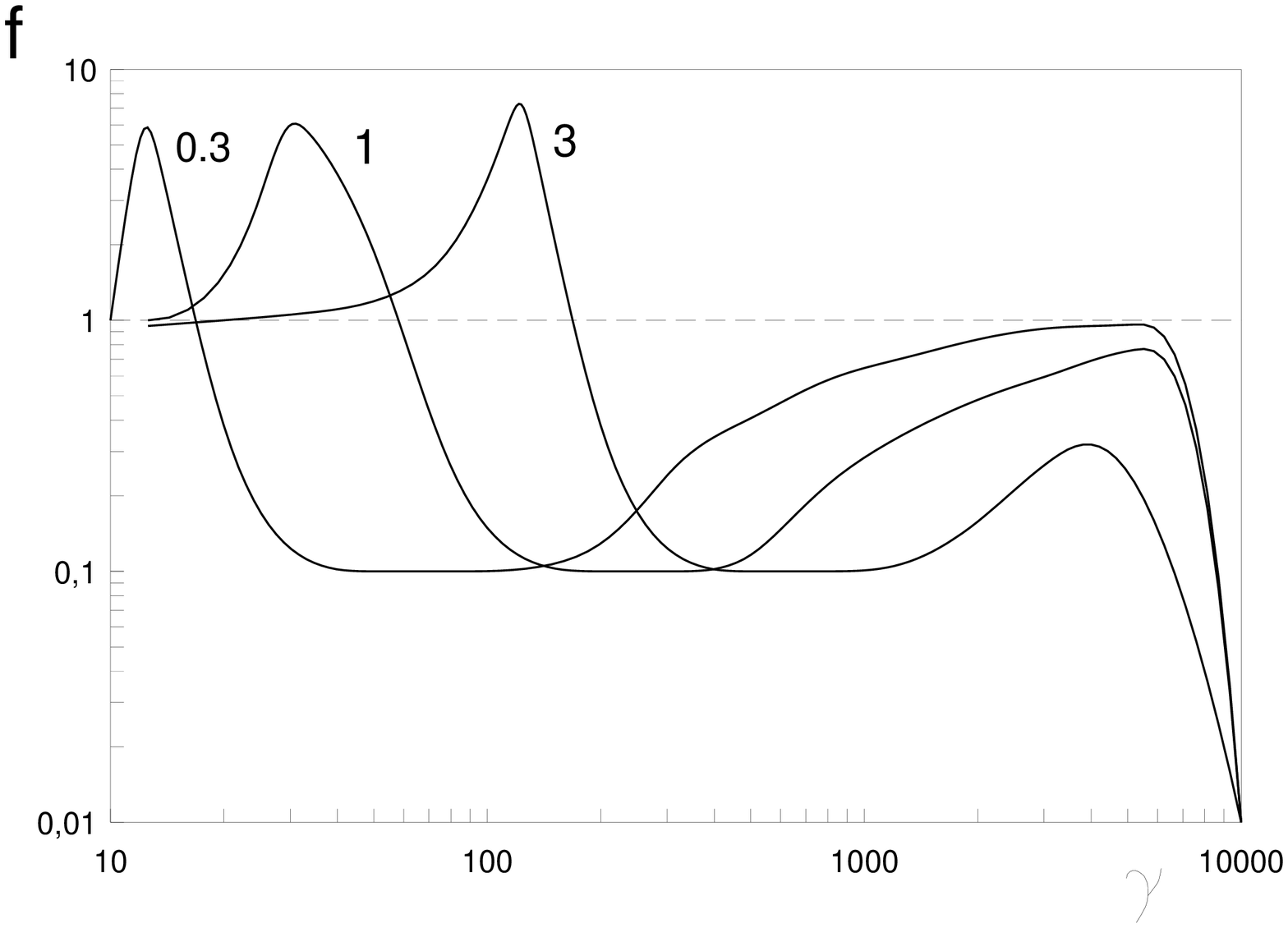}
\caption{
 Evolution of the waterbag distribution function on account of
resonant inverse Compton scattering: {\bf a} for various temperatures,
$T_6$; $B_{12}=1$, {\bf b} for various magnetic field strengths,
$B_{12}$; $T_6=0.5$; the initial distribution function is shown by the dashed
line
}
\end{figure}

\begin{figure}[htb]
\includegraphics[scale=0.4]{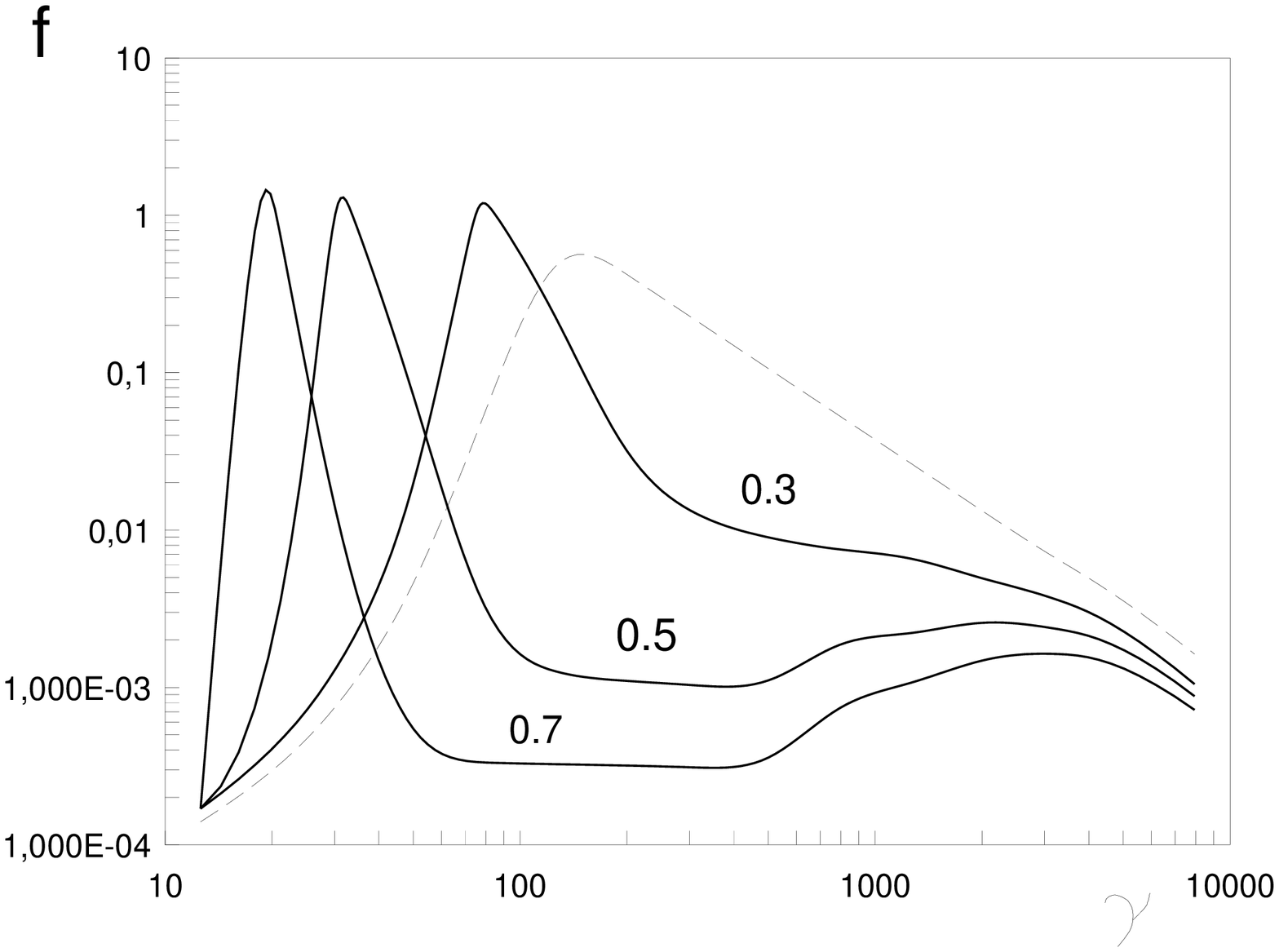}
\includegraphics[scale=0.4]{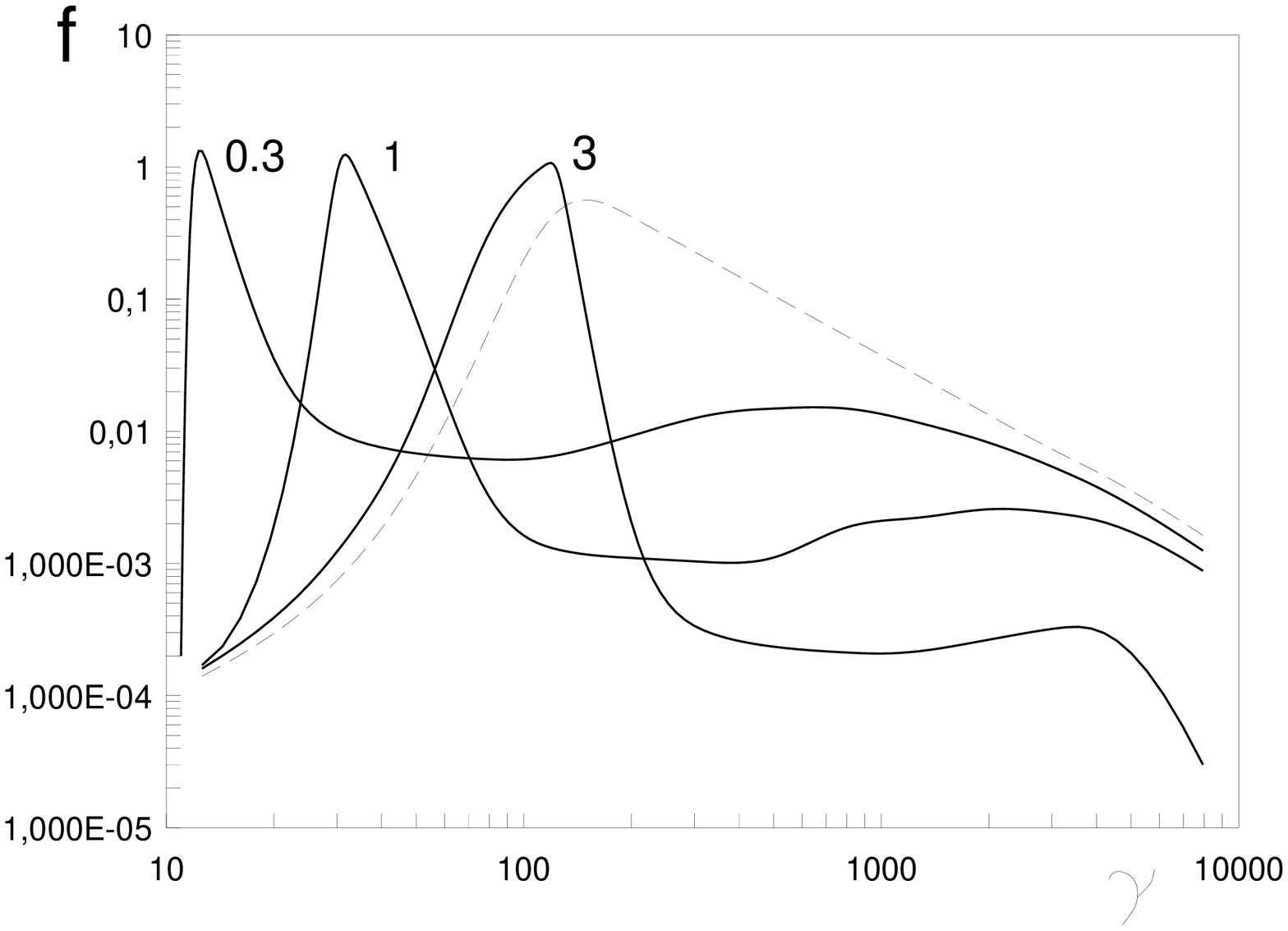}
\caption{
 The same as in Fig. 5 for the initial distribution function (5).
}
\end{figure}

\end{document}